\documentclass[runningheads]{llncs}
\usepackage[hyperref=auto,style=numeric-comp,url=false]{biblatex}
\bibliography{kbib/kwarc,mkm2010}
\usepackage{lststex}
\usepackage{lstgrgen}
\usepackage{stex-logo}
\usepackage[show]{ed}
\usepackage{paralist}
\usepackage{wrapfig}
\usepackage{xcolor}
\usepackage{graphicx}
\usepackage{xspace}
\usepackage{paralist}
\usepackage{tikz}
\usetikzlibrary{shapes,snakes}
\usetikzlibrary{positioning}
\usetikzlibrary{fit}
\usepackage{pgf}

\def\sys{\textsc{Planetary}\xspace} 
\newcommand{\doctip}{\textsl{DocTIP}\xspace}
\newcommand{\grgen}{\textsl{GrGen.NET}\xspace}
\newcommand{\gmoc}{\textsl{GMoC}\xspace}
\newcommand{\tntbase}{\textsl{TNTBase}\xspace}
\newcommand{\doors}{\textsl{Doors}\xspace}
\newcommand{\OMDOC}{OMDoc\xspace}

\title{Workflows for the Management of Change in Science,
  Technologies, Engineering and Mathematics}

\titlerunning{Workflows for the Management of Change in STEM}
\author{Serge Autexier\inst{1} \and 
  Catalin David \inst{1,2} \and 
  Dominik Dietrich\inst{1} \and 
  Michael Kohlhase\inst{2} \and 
  Vyacheslav Zholudev\inst{1,2}}
\authorrunning{Autexier, David, Dietrich, Kohlhase, Zholudev}
\institute{
\begin{tabular}[t]{c}
  Safe and Secure Cognitive Systems, German Research Centre for\\
  Artificial Intelligence (DFKI), Bremen, Germany
\end{tabular}
\and
\begin{tabular}[t]{c}
  Computer Science, Jacobs University Bremen, Germany
\end{tabular}
}

\begin{document}
\maketitle

\begin{abstract}
Mathematical knowledge is a central component in science,
 engineering, and technology (documentation). Most of it is
  represented informally, and -- in contrast to published research
  mathematics -- subject to continual change. Unfortunately, machine
  support for change management has either been very coarse grained
  and thus barely useful, or restricted to formal languages, where
  automation is possible.
  In this paper, we report on an effort to extend change management to
  collections of semi-formal documents which flexibly intermix
  mathematical formulas and natural language and to integrate it into
  a semantic publishing system for mathematical knowledge. We validate
  the long-standing assumption that the semantic annotations in these
  flexiformal documents that drive the machine-supported interaction
  with documents can support semantic impact analyses at the same
  time. But in contrast to the fully formal setting, where adaptations
  of impacted documents can be automated to some degree, the
  flexiformal setting requires much more user interaction and thus a
  much tighter integration into document management workflows.
\end{abstract}

\section{Introduction}\label{sec:intro}

As the Web 2.0 age is dawning for mathematics, more and more
{\emph{mathematical development}} is moving online; not just
\emph{publications}. An example of this is the PolyMath site, where
upon the recent announcement of a proof of $P\ne NP$, the mathematics
community has organized itself in a WiKi and found a significant gap
in the proof within two weeks; see~\cite{PolymathPvsNP}. The
PlanetMath community which has collaborated on 8500 graduate-level
encyclopedia articles over 10 years~\cite{planetmath:on} is another,
and also the Mizar community, who have formalized more than 60000
definitions, assertions, and proofs and have machine-checked them over
the last 40 years. Finally, the Cornell EPrint
Archive~\cite{arxiv:online} has amassed over 660 000 scientific
articles over 20 years. The hallmark of all these efforts is that they
are massive collaborations by many individuals, distributed widely
both geographically and temporally. The first three examples have
another characteristic that is becoming more and more important: the
knowledge items are interdependent and mutable (subject to
change). The sheer size of the knowledge collections together with the
fact that many authors do not even know (of) each other induces
consistency and coherence problems. In this situation, the need to
integrate the mechanisms for ``change management'' (CM) into the
digital libraries seems obvious. Typically, the documents in the
libraries are \emph{flexiformal} (flexibly formal) because they
contain semantic annotations at different levels of formality. A good
example is an informal, but rigorous statement from a mathematical
textbook, which intermixes mathematical formulas (formal
representations of mathematical objects) with natural language
(informal representations of their relations).  Change management makes use of the
fact that MKM formats explicitly represent the relations between
objects to compute related objects and predict the way changes affect
them; see~\cite{Hutter:smhd09,NMueller:PhD,AM-10-a} for recent
progress in this field.

This paper reports on the experiment of integrating CM into the \sys system, a new
flexiformal Digital library system, which we will present in the next section. In
Section~\ref{sec:workflow}, we describe the information present in the sources by way of an
extended example and show how these can be used for change management. In
Section~\ref{sec:doctip}, we present the \doctip system and the CM procedure it
implements, so that we can show the integration from an architectural point of view in
Section~\ref{sec:architecture}. Section~\ref{sec:reex} revisits the example from
Section~\ref{sec:workflow} to show how the information travels through the systems
involved. In Section~\ref{sec:related}, we discuss related work and Section~\ref{sec:concl}
concludes the paper.

\section{The \sys System}\label{sec:planetary}
The \sys system (see~\cite{KohDavGin:psewads11,DGKC:eMath30,Planetary:on} for an
introduction) is a Web 3.0 system\footnote{We adopt the nomenclature where Web 3.0 stands
  for extension of the Social Web with Semantic Web/Linked Open Data technologies.} for
semantically annotated document collections in Science, Technology, Engineering and
Mathematics (STEM).  The system is based on {\emph{semantically annotated documents}}
together with semantic background ontologies (which we call the {\bf{content
    commons}}). This information can then be used by user-visible, semantic services like
program (fragment) execution, computation, visualization, navigation, information
aggregation and information retrieval. Finally a document player application can embed
these services to make documents executable. We call this framework the {\textbf{Active
    Documents Paradigm}} (ADP), since documents can also actively adapt to user
preferences and environment rather than only executing services upon user request.

\begin{figure}[t]
\centering\includegraphics[width=12cm]{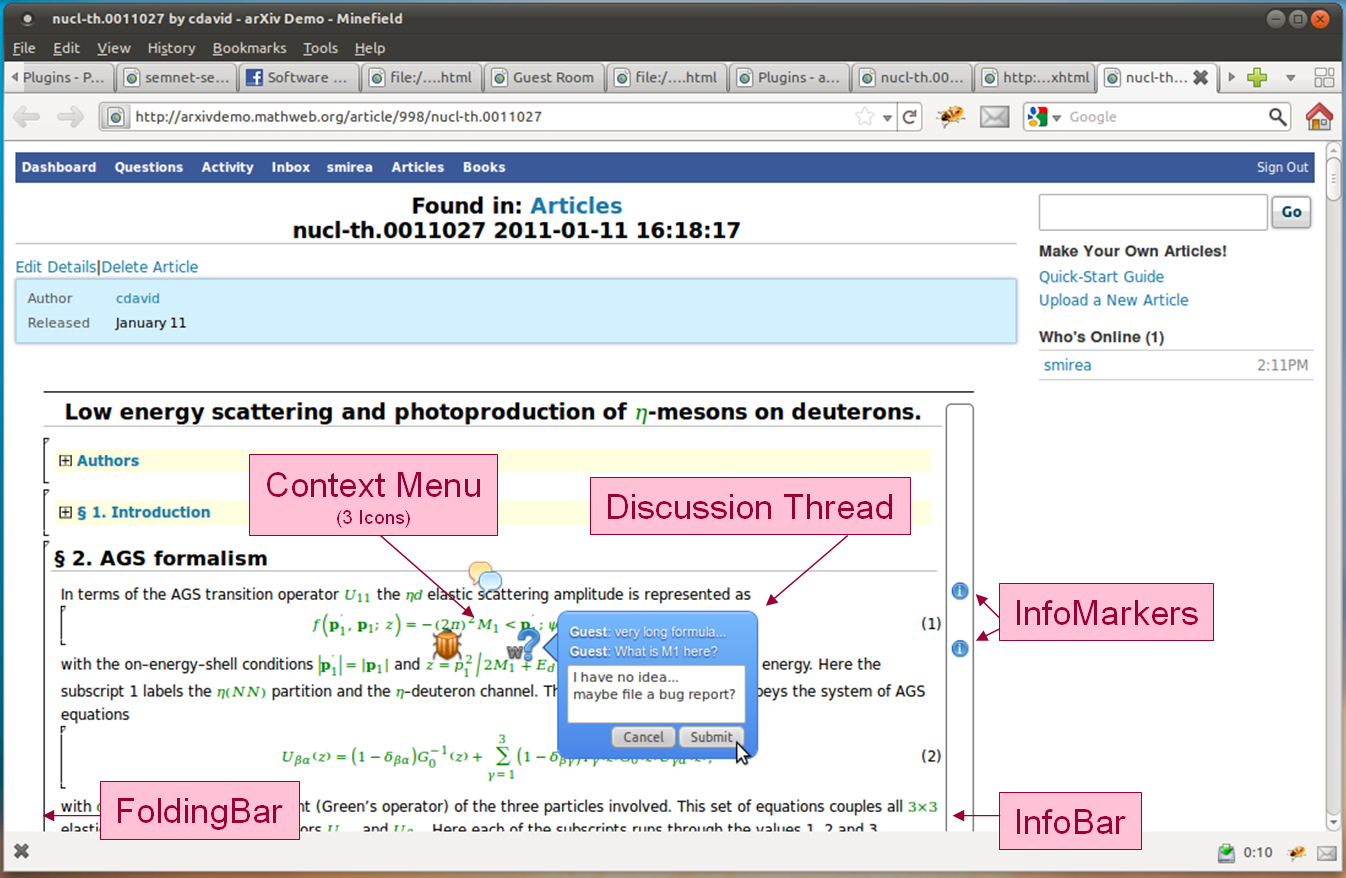}
\caption{Course Notes in the \sys system}\label{fig:planetary}
\end{figure}

In our approach, {\emph{documents published in the \sys system become flexible, adaptive
    interfaces to a content commons}} of domain objects, context, and their relations.
The system achieves this by providing embedded user assistance through an extended set of
user interactions with documents based on an extensible set of client- and server side
services that draw on explicit (and thus machine-understandable) representations in the
content commons (see Fig.~\ref{fig:planetary}). 

The \sys system has been used on the course notes of a two-semester introductory course in
Computer Science~\cite{Kohlhase:PlanetGenCS:url} held at Jacobs University by one of the
authors in the last eight years. While the basic concept of the course stayed the same
over the years, whole topics have been added/moved/deleted, examples and results have been
added, and formulations have been sharpened. All of these changes had consequences that
were sometimes difficult to foresee, and sometimes led to problematic teaching situations
(when the consequences had not been anticipated). The course notes currently comprise 300
pages with over 500 slides organized in over 800 files. This is at the limits of what is
manually manageable for the instructor who has authored all of the material; it would be
impossible for a new instructor to take over the material (and change it to her
liking). It becomes increasingly difficult to manage the over 1000 homework, quiz, and
exam problems that have largely been provided by the more than 30 teaching assistants that have
accompanied the course over the years.

\section{A \sys Workflow}\label{sec:workflow}

\begin{figure}[t]
\centering
\begin{tabular}{|l|}\hline
\begin{lstlisting}
\begin{module}[id=binary-trees]
  \importmodule[\KWARCslides{graphs-trees/en/trees}]{trees}
  \importmodule[\KWARCslides{graphs-trees/en/graph-depth}]{graph-depth}
  ...
  \begin{definition}[id=binary-tree.def,title=Binary Tree]
    A \definiendum[binary-tree]{binary tree} is a \termref[cd=trees,name=tree]{tree}
    where all \termref[cd=graphs-intro,name=node]{nodes}
    have \termref[cd=graphs-intro,name=out-degree]{out-degree} 2 or 0.
  \end{definition}
  ...
  \begin{definition}[id=bbt.def]
    A \termref[name=binary-tree]{binary tree} $G$ is called 
    \definiendum[bbt]{balanced binary tree} iff the 
    \termref[cd=graph-depth,name=vertex-depth]{depth} of all
    \termref[cd=trees,name=leaf]{leaves} differs by at most by 1, and
    \definiendum[fullbbt]{fully balanced}, iff the
    \termref[cd=graph-depth,name=vertex-depth]{depth} difference is 0.
  \end{definition}
  ...
\end{module}
\end{lstlisting}
\\\hline
\includegraphics[width=12cm]{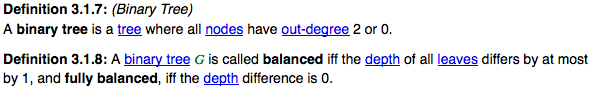}
\\\hline
\end{tabular}
\caption{Two definitions and their \protect\stex sources}\label{fig:defs}
\end{figure}

To get a better intuition for the problems involved in managing changes in flexiformal
document collections, consider the situation in Fig.~\ref{fig:defs} and~Fig.~\ref{fig:pf},
which we will use as a running example. The lower part of Fig.~\ref{fig:defs} shows two
well-known definitions from the theory of binary trees and Fig.~\ref{fig:pf} a lemma
that depends on them, as they are referenced in its proof.  Clearly, if one of the
definitions is changed, then we have to revisit the proof and possibly adapt it or
even the lemma to the changed situation.

For humans, it is simple to detect the underlying dependency in principle, but there is a
strong possibility that it will be overlooked in practice; especially, if the conceptional
distance between a proof and the definitions is large (e.g., because it involves many
intervening definitions and assertions). Therefore, authors need system support to keep
large mutable knowledge collections in a consistent state. In the situation of our running
example, we can make use of the fact that the two text fragments were originally
written as semantically annotated \stex course notes~\cite{Kohlhase:PlanetGenCS:url} for
\sys. As such, they contain a lot of semantic annotations that are originally added to
drive services like definition lookup, notation adaptation, and just-in-time prerequisites
delivery, which also induce a good approximation of the semantic dependency relation that
is needed for analysing the impact of changes on definitions and proofs in this and other knowledge items. 

\begin{figure}[t]
\centering
\begin{tabular}{|p{12cm}|}\hline
\begin{lstlisting}[aboveskip=-1ex]
\begin{module}[id=bbt-size]
\importmodule[binary-trees]{binary-trees}
\end{lstlisting}
\ldots\\\hline
\quad\begin{minipage}{11cm}
{\textbf{Lemma 3.1.9} \it  Let $G=\langle V,E\rangle$ be a \underline{balanced binary tree} of \underline{depth} $n>i$, then the set
    $V_i:=\{v\in V:dp(v=i)\}$ of \underline{vertexes} at \underline{depth} $i$ has \underline{cardinality} $2^i$.}\\
  \textbf{Proof:} by \underline{induction} over the \underline{depth} $i$
\end{minipage}\\\hline
\begin{lstlisting}[aboveskip=-1ex,belowskip=-2ex]
   \begin{spfstep}
      By the \begin{justification}[method=byDef]
        \premise[uri=binary-trees,ref=binary-tree.def]{definition of a binary tree}
      \end{justification}, each $\inset{v}{V_{i-1}}$ is a leaf or has
      two children that are at depth $i$.
    \end{spfstep}
    \begin{spfstep}
       As $G$ is \termref[cd=binary-trees,name=bbt]{balanced} 
       and $\gdepth{G}=n>i$, $V_{i-1}$ cannot contain leaves.
     \end{spfstep}
     ...
 \end{sproof}
\end{module}
\end{lstlisting}
\\\hline
\end{tabular}
\caption{A lemma and proof that depend on the definitions in Fig.~\ref{fig:defs}}\label{fig:pf}
\end{figure}
Let us consider these annotations in the \stex sources in Fig.~\ref{fig:defs}
and~Fig.~\ref{fig:pf}.  In the first proof step (the \stex \lstinline{spfstep} environment) in
Fig.~\ref{fig:pf}, the ``definition of a binary tree'' is referenced, and this reference
is marked up by a URI reference encoded in the optional argument of the
\lstinline{premise} macro inside the \lstinline{justification} element. In the second
proof step, the property of being ``balanced'' is exploited. The fact that
the word ``balanced'' is used as a technical term is marked up with the
\lstinline{\termref} macro, whose optional first argument points to the
\lstinline{\definiendum} with name \lstinline{bbt} in the module
{\lstinline{binary-trees}} in Fig.~\ref{fig:defs}.

Intuitively, the relations encoded in these annotations induce the dependency that signals
a possible semantic impact of a change to one of the definitions in
Fig.~\ref{fig:defs}. There are at least three possible ways an author can benefit from
an automated impact analysis based on the semantic annotations in the \stex sources.
\begin{enumerate}[\bf C1]
\item An author who wants to change something in one (or both) of the definitions in
  Fig.~\ref{fig:defs} can request an estimation of the total impacts costs of a
  change.
\item An author who actually changes (one of) the definitions can
  request an immediate impact analysis, which gives a list of
  potentially affected knowledge items. This list should be
  cross-linked to the (presentations of) the affected items, so to
  simplify navigation. For every item the author will have to decide whether
  it is really affected and how to adapt it (possibly creating new
  impacts in the process).
\item Authors or maintainers of a given knowledge item can be notified of an impact to
  ``their'' knowledge item upon changes to elements it depends on. 
\end{enumerate}
Note that \textbf{C1} and \textbf{C2} together constitute what one could call a ``push
workflow of change management'' whereas \textbf{C3} corresponds to a ``pull workflow''. The
abundance of semantic references --- 12 in this little example --- already shows that
machine support is indispensable in larger collections.  Note furthermore that both of
these workflows should be completely independent of the ``commit policies'' of the
knowledge collection. The change management subsystem should support committing partially
worked off impact lists --- e.g., for the weekend or to pass them on to other authors.

\section{\doctip}\label{sec:doctip}
The \doctip system~\cite{doctip:online} provides a generic framework that combines
sophisticated structuring mechanisms for heterogenous formal and
semi-formal documents with an appropriate change management to
maintain structured relations between different documents. It is based
on abstract document models and abstract document ontologies that need
to be instantiated for specific document kinds, such as \OMDOC. The
heart of the system is the \emph{document broker}, which maintains all
documents and provides a generic update and patch-based
synchronisation protocol between the maintained documents and the
connected \emph{components} working on these documents. Components can
be authoring (and display) systems, or analysis and reasoning systems
offering automatic background processing support, or simply a
connection to a repository allowing to commit and update the
documents.

If the document broker obtains a change for some of its documents, the
changes are propagated to all connected components for that
document. A configurable impact analysis policy allows the system
designer to define if impact analysis is required after obtaining a
change from some component. To perform the impact analysis the
document broker uses the \gmoc\footnote{\gmoc: \underline{G}eneric \underline{M}anagement \underline{o}f \underline{C}hange} tool (\cite{AM-10-a} see below) to
compute the effect of the change on all documents maintained by the
document broker. The \gmoc tool returns that information as impact
annotations to each individual document, which are subsequently
distributed to all connected components by the document broker.

\subsection{Change Impact Analysis}\label{sec:cia}
The key idea to design change impact analysis (CIA) for informal
documents is the \emph{explicit semantics method} which represents
both the syntax parts (i.e., the documents) and the intentional
semantics contained in the documents in a single, typed hyper-graph
(see \cite{AM-10-a} for details). Document type specific graph
rewriting rules are used to extract the intentional semantics of
documents and the extracted semantic entities are linked to their
syntax source, i.e. their \emph{origin}. That way, any change in the
document results in semantic objects for which origins have been
deleted or changed, as well as syntax objects for which there does not
exist corresponding semantic entities yet. The semantic objects are
marked with this status information (``deleted'', ``added'',
``preserved'').  This information is then exploited by analysis rules
to compute the ripple effects of the changes on the semantics
entities, which in a final stage are used to annotate the syntax
parts, that is the documents. The \gmoc tool is built on top of the
graph rewriting tool \grgen~\cite{grgen} and is parameterized over
document type specific document meta-models and graph rewriting rule
systems to extract the semantics and to analyze the impact of
changes. \vspace*{1ex}

\paragraph{Document Meta-Models.}\label{sec:metamodels}
To provide change impact analysis for \sys, we developed a document meta model and
graph impact analysis rules for \OMDOC. The document meta model consists of a lightweight
ontology of the relevant semantic concepts in \OMDOC documents, --- e.g., theories, symbol
declarations and their occurrences, axioms, definitions, assertions, and their use in
proofs and proof steps --- together with semantic relations between concepts ---
e.g., import relations between theories, symbols and their definitions, assertions and
their proofs. Note that the \OMDOC meta-model abstracts over the \OMDOC surface
syntax. For instance, a definition can either be a
\lstinline[language=XML]{definition}-element \vspace*{2ex}

\begin{lstlisting}[language=XML,mathescape,basicstyle=\footnotesize\sf]
<symbol name=''unit''>
<definition xml:id="mon-d1" for="unit" type="informal">
 <CMP> 
  A structure $\fbox{(M,*,e)}$, in which $\fbox{(M,*)}$ is a semi-group with unit $\fbox{e}$ is called monoid.
 </CMP>
</definition>
\end{lstlisting}\vspace*{2ex}
\noindent where the \lstinline[language=XML]{symbol} defined by the definition is given by the
\lstinline[language=XML]{for} attribute of the definition (boxes abbreviate OpenMath
content here). The \lstinline[language=XML]{symbol} itself is declared in a different
element. This kind of definition typically occurs when \OMDOC documents are created
manually or obtained from formal representations.  Alternatively, a definition can come as
a ``typed'' \lstinline[language=XML]{omtext} such as \vspace*{2ex}

\begin{lstlisting}[language=XML,basicstyle=\footnotesize\sf,showspaces=no,mathescape]
<omtext type="definition" xml:id="binary-tree.def" about="#binary-tree.def">
 <CMP xml:id="binary-tree.def.CMP1" about="#binary-tree.def.CMP1">
  <p xml:id="binary-tree.def.CMP1.p1" about="#binary-tree.def.CMP1.p1">
   A <term cd="balanced-binary-trees" name="binary-tree" role="definiendum">
   binary tree</term> is a <term cd="trees" name="tree" 
   xml:id="binary-tree.def.CMP2.p1.term2" 
   about="#binary-tree.def.CMP2.p1.term2">tree</term> where all $\ldots$
  </p>
 </CMP>
</omtext>
\end{lstlisting}\vspace*{2ex}
\noindent 
which typically happens, for instance, when generating the \OMDOC files from an \stex
source file. Note that in this case the defined \lstinline[language=XML]{symbol} is
declared by the \lstinline[language=XML]{term} element with
\lstinline[language=XML]{role="definiendum"}. The fact that this definition defines that
symbol comes from the structural nesting of the term inside the definition. Similar
examples are theories which can either be imported into each other by using the explicit
\lstinline[language=XML]{imports} elements or simply by nesting
\lstinline[language=XML]{theory}-elements.

Conceptually, it does and should not matter in which form symbols and
definitions are given, and a mixture of both forms is also desirable
to support the linking of mathematical content in \OMDOC from
different authoring sources. The document meta model declares these
pure concepts and relations like an ontology. The intentional
semantics of a given \OMDOC document is a set of instances of these
concepts and relations. The used graph rewriting tool supports
hypergraphs with typed nodes and edges. The types are simple types
with sub-typing relations. This is exploited to subdivide the whole
graph in a syntax and a semantic subgraph by introducing top-level
types for either part.  The \OMDOC syntax elements are declared as
subtypes of the syntax type and the \OMDOC document being an XML tree
can then naturally be represented as (syntax) nodes and
relations. Analogously, the semantic concepts and relations from the
\OMDOC document meta-model are simply declared as subtypes of the
semantic types.

\tikzstyle{impact} = [ellipse, draw, fill=blue!20,minimum width=.5cm, text centered, rounded corners, minimum height=1cm,shade, top color=white, bottom color=blue!20]
\tikzstyle{block} = [rounded rectangle, draw, fill=blue!20, minimum width=.5cm, text centered, rounded corners, minimum height=1cm,shade, top color=white, bottom color=blue!20 ]
\tikzstyle{textnode} = [rectangle, draw, fill=blue!20, minimum width=.5cm, text centered, minimum height=1cm]

\tikzstyle{negblock} = [block,dashed, bottom color=white] 
\tikzstyle{negimpact} = [impact,dashed, bottom color=white]
\tikzstyle{negtextnode} = [textnode,dashed,  bottom color=white]

\tikzstyle{pac} = [rectangle split, rectangle split parts=2, draw, text width=5cm, text centered, rounded corners, minimum height=1cm]
\tikzstyle{fitbox} = [rectangle, draw, inner sep=0.5cm]
\tikzstyle{negline} = [->,dashed]
\tikzstyle{posline} = [->,draw]

\def\scalefactor{.5}

\begin{figure}[t]
  \centering
  \begin{tabular}{ccc}


\begin{tikzpicture}[scale=\scalefactor,transform shape] 

\node[rectangle,draw,minimum height=1.2cm] (lhs){
\begin{tikzpicture}[node distance = 1cm]
    \node [textnode] (omtext) {omtext};
    \node [right] (labelomtext) at (omtext.east) {d};

    \node [textnode,below=of omtext] (box1) {Attribut type};
    \node [right] (labelbox1) at (box1.east) {p};

    \node [textnode, left=1.5cm of box1] (box2) {Attribut xml:id};
    \node [right] (labelbox2) at (box2.east) {x};
    
    \node [negblock, left=1.5cm of omtext] (negbox) {OMDoc Definition};
      
  \draw[posline] (box1) -- node[midway,right] (labelarrow) {isAttribute} (omtext);
  \draw[posline] (box2) -- (omtext);    
  \draw[negline] (omtext) -- node[midway,above] (labelarrow) {origin} (negbox);
\end{tikzpicture}
};
\node[rectangle, anchor=north west, draw] at (lhs.north west) {L};

\node [below=2cm of lhs,rectangle split, rectangle split parts = 2,draw] (rhs) {

\begin{tikzpicture}[node distance = 1cm]
    \node [textnode] (omtext) {omtext};
    \node[above=0cm of omtext, minimum height=0cm, inner sep=2pt] {};
    \node [right] (labelomtext) at (omtext.east) {d};

    \node [textnode, below=of omtext] (box1) {Attribut type};
    \node [right] (labelbox1) at (box1.east) {p};

    \node [textnode, left=1.5cm of box1] (box2) {Attribut xml:id};
    \node [right] (labelbox2) at (box2.east) {x};

	\node [block, left=1.5cm of omtext] (box3) {Omdoc Definition};
    \node [below] (label2) at (box3.south) {od};
      
  \draw[posline] (box1) -- node[midway,right] (labelarrow) {isAttribute} (omtext);
  \draw[posline] (box2) -- (omtext);
  \draw[posline] (omtext) -- node[midway,above] (labelarrow) {origin} (box3);
\end{tikzpicture}
\nodepart{second}
\begin{tikzpicture}[node distance=1mm]
  \node[inner sep=0mm] (status1) {od.status = added}; 
  \node[inner sep=0mm,below=of status1] (status2)  {od.xmlid = x.value};
  \node[inner sep=0mm,below=of status2] (status3)  {detectOMTextDefiniendum(d,od)};
  \node[inner sep=0mm,below=of status3] (status4)  {detectCMP(od)};
\end{tikzpicture}
};
\node[rectangle, anchor=north west, draw] at (rhs.north west) {R};
\node[rectangle, anchor=north west, draw] at (rhs.text split west) {Apply};

\node[below=0.1cm of lhs] (lhs2) {};
\node[above=0.1cm of rhs] (rhs2) {};
\draw[->, line width=3pt] (lhs2) -- (rhs2);

\end{tikzpicture}
    & 


\begin{tikzpicture}[scale=\scalefactor,transform shape] 

\node[rectangle,draw,minimum height=1.2cm,rectangle split, rectangle split parts = 2] (lhs){
\begin{tikzpicture}[node distance = 1cm]
    \node [textnode] (omtext) {omtext};
    \node [right] (labelomtext) at (omtext.east) {d};

    \node [textnode, below=of omtext] (box1) {Attribut type};
    \node [right] (labelbox1) at (box1.east) {p};

    \node [textnode, left=1.5cm of box1] (box2) {Attribut xml:id};
    \node [right] (labelbox2) at (box2.east) {x};

	\node [block, left=1.5cm of omtext] (negbox) {OMDoc Definition};
  \node [below] (labelomtext) at (negbox.south) {od};
      
  \draw[posline] (box1) -- node[midway,right] (labelarrow) {isAttribute} (omtext);
  \draw[posline] (box2) -- (omtext);    
  \draw[posline] (omtext) -- node[above,midway] (labelarrow) {origin} (negbox);
\end{tikzpicture}
\nodepart{second}
\begin{tikzpicture}[node distance = 0cm]
    \node [rectangle] (status1) {od.status == deleted};
\end{tikzpicture}
};
\node[rectangle, anchor=north west, draw] at (lhs.north west) {L};
\node[rectangle, anchor=north west, draw] at (lhs.text split west) {PAC};

\node [below=2cm of lhs,rectangle split, rectangle split parts = 2,draw] (rhs) {

\begin{tikzpicture}[node distance = 1cm]
    \node [textnode] (omtext) {omtext};
    \node[above=0cm of omtext, minimum height=0cm, inner sep=2pt] {};
    \node [right] (labelomtext) at (omtext.east) {d};

    \node [textnode, below=of omtext] (box1) {Attribut type};
    \node [right] (labelbox1) at (box1.east) {p};

    \node [textnode, left=1.5cm of box1] (box2) {Attribut xml:id};
    \node [right] (labelbox2) at (box2.east) {x};

	\node [block, left=1.5cm of omtext] (box3) {Omdoc Definition};
    \node [below] (labelomtext) at (box3.south) {od};
      
  \draw[posline] (box1) -- node[right,midway] (labelarrow) {isAttribute} (omtext);
  \draw[posline] (box2) -- (omtext);
  \draw[posline] (omtext) -- node[midway,above] (labelarrow) {origin} (box3);
\end{tikzpicture}
\nodepart{second}
\begin{tikzpicture}[node distance = 1mm]
    \node [rectangle] (status1) {od.status=preserved};
  \node [rectangle,inner sep=0mm,below=of status1] (status3) {detectOMTextDefiniendum(d,od)};
  \node [rectangle,inner sep=0mm,below=of status3] (status4) {detectCMP(od)};
\end{tikzpicture}
};
\node[rectangle, anchor=north west, draw] at (rhs.north west) {R};
\node[rectangle, anchor=north west, draw] at (rhs.text split west) {Apply};

\node[below=0.1cm of lhs] (lhs2) {};
\node[above=0.1cm of rhs] (rhs2) {};
\draw[->, line width=3pt] (lhs2) -- (rhs2);

\end{tikzpicture}
    &


\begin{tikzpicture}[scale=\scalefactor,transform shape] 

\node[rectangle,draw,minimum height=1.2cm] (lhs){
\begin{tikzpicture}[node distance = 1cm]
    \node [block] (omdef) {Omdoc Definition};
    \node[above=0cm of omdef, minimum height=0cm, inner sep=2pt] {};
    \node [right] (labelomtext) at (omdef.north east) {$\quad$od};

    \node [block, below=of omdef] (box2) {TheoryObject};
    \node [right] (labelbox2) at (box2.north east) {$\quad$imp};

    \node [impact,right=of omdef] (defchanged2) {Definition changed};
    \node [right] (labelbox1) at (defchanged2.east) {a};

    \node [negimpact, right=of box2] (negbox) {Definition changed};
      
    \draw[posline] (defchanged2) -- node[midway,right] (labelarrow) {} (omdef);
    \draw[posline] (box2) -- node[midway, right=0.1cm] (labelarrow) {uses} (omdef);    
    \draw[negline] (box2) -- (negbox);
\end{tikzpicture}
};
\node[rectangle, anchor=north west, draw] at (lhs.north west) {L};

\node [below=2cm of lhs,rectangle split, rectangle split parts = 2,draw] (rhs) {

\begin{tikzpicture}[node distance = 1cm]
    \node [block] (omdef) {Omdoc Definition};
    \node[above=0cm of omdef, minimum height=0cm, inner sep=2pt] {};
    \node [right] (labelomtext) at (omdef.east) {od};

    \node [block, below=of omdef] (box2) {TheoryObject};
    \node [right] (labelbox2) at (box2.east) {imp};

    \node [impact,right=of omdef] (defchanged2) {Definition changed};
    \node [right] (labelbox1) at (defchanged2.east) {a};

    \node [impact, right=of box2] (negbox) {Definition changed};
    \node [below] (labelomtext) at (negbox.south) {l};
      
    \draw[posline] (defchanged2) -- node[midway,right] (labelarrow) {} (omdef);
    \draw[posline] (box2) -- node[midway, right=0.1cm] (labelarrow) {uses} (omdef);    
    \draw[posline] (box2) -- (negbox);
\end{tikzpicture}
\nodepart{second}
\begin{tikzpicture}[node distance = 1mm]
    \node [rectangle] (status1) { $ $};
  \node [rectangle,inner sep=0mm,below=0.3cm of status1] (status3) {a.desc="Used target definition changed"};
  \node [rectangle,inner sep=0mm,below=of status3] (status4) {emit "Used target definition changed"};
\end{tikzpicture}
};
\node[rectangle, anchor=north west, draw] at (rhs.north west) {R};
\node[rectangle, anchor=north west, draw] at (rhs.text split west) {Apply};

\node[below=0.1cm of lhs] (lhs2) {};
\node[above=0.1cm of rhs] (rhs2) {};
\draw[->, line width=3pt] (lhs2) -- (rhs2);

\end{tikzpicture}\\
    \texttt{FindNewDefinition}
    &
    \texttt{FindExistingDefinition}
    & 
    \texttt{PropagateChangedDefinition}
  \end{tabular}
  \caption{Two Abstraction Rules and one Propagation Rule written
    top-down; we use rectangles for syntax nodes, rounded rectangles
    for semantic nodes and ellipses for impact nodes.}
  \label{fig:sampelabstractionrule}
  \label{fig:rulepropagate}
\end{figure}
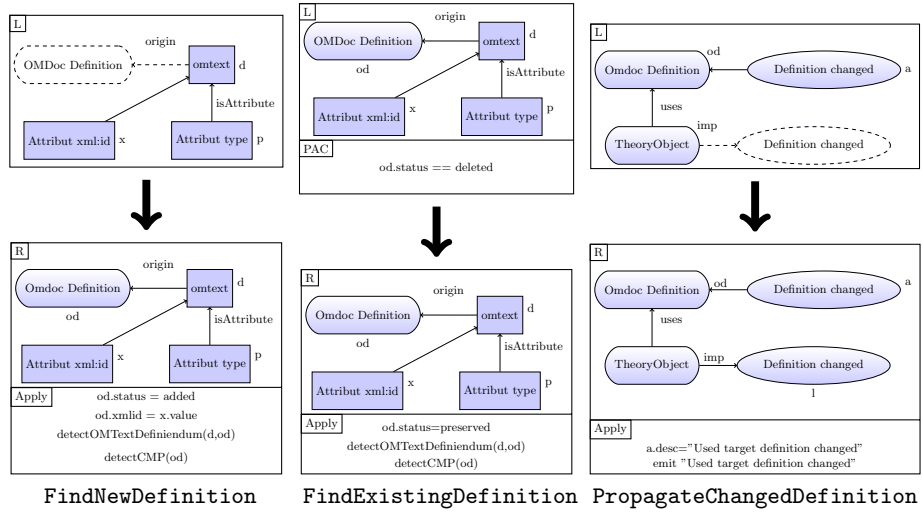

\paragraph{Abstraction Phase of CIA.}\label{label:sec:cia:abstraction}
The abstraction phase of the impact analysis for \OMDOC documents
consists of extracting the intentional semantics from the given \OMDOC
documents. This is realized by a set of graph rewriting rules which
analyse the \OMDOC document to extract the semantic concepts and
relations, and mark them as being added. Examples of such rules are
the two left-most rules in Fig.~\ref{fig:sampelabstractionrule} to
extract definitions from ``typed'' \lstinline[language=XML]{omtext}:
The graph rewriting rules are named (e.g., \texttt{FindNewDefinition})
and have a left-hand side (the box labelled by \texttt{L}) indicating
the pattern to match in a subgraph and a right-hand side (the box
labelled \texttt{R}) by what the instantiated subgraph pattern is
replaced. Identical graph nodes and edges are additionally labelled by
names, such as $x,d,p,od, \ldots$. Further conditions that must be
satisfied to enable the graph rewriting step are positive application
conditions (PAC), which must hold on the graph before rule application
and negative application conditions (the dashed nodes and edges in the
left-hand sides \texttt{L} or in extra NAC boxes---not
used here) which must be false on the graph before rule
application. These conditions can be graph patterns as well as boolean
tests on attribute values. The application of the graph rewriting rule
replaces the subgraph in \texttt{L} with the subgraph in \texttt{R}
and additional adaptations can be triggered in the \texttt{Apply}
part, such as adapting the value of attributes but also invoking
further graph rewriting rules using their name (e.g.,
\texttt{detectCMP}).

The rules for the abstraction phase always come in two variants: one
variant is for new syntactic \lstinline[language=XML]{omtext}s, i.e.,
there does not exist yet a semantic object in the semantic
graph. For these, new semantic instances are introduced, marked as
added and the origin of the semantic concept is represented explicitly
by an \lstinline[language=GRG]{Origin} edge from the semantic node to
the syntax node. The second variant is for already known syntactic
\lstinline[language=XML]{omtext}s, i.e., there exist already a
semantic object in the semantic graph from a previous version of the
document. For these, the semantic instances are maintained and marked
as preserved.  Both rules invoke further rules to analyse the
``body'' of a definition in order to find out whether the definition
has changed (e.g., \lstinline[language=GRG]{detectCMP}). All semantic
objects that are neither added nor preserved are marked as deleted by
a generic rule operating over all semantic nodes and edges. Overall we
have designed 91 rules for the abstraction phase that synchronizes
\OMDOC documents with their intentional semantics.

\begin{figure}[t]\centering\vspace*{-3mm}\def\meta#1{\ensuremath{\langle\mbox{\sl #1}\rangle}}
\begin{lstlisting}[language=XML, basicstyle=\small,mathescape]
<impacts>
  <impact for="binary-tree.def" name="Definition changed" 
          select="$\meta{xpath-to-definition-binary-tree}$"/>
  <impact for="balanced-binary-tree.def" name="Definition Binary Tree changed" 
          select="$\meta{path-to-definition-balanced-binary-tree}$"/>
  <impact for="size-lemma-pf.derive2.method2.proof2.derive4.CMP1.p1.term2"
          name="Definition Binary Tree changed" 
          select="$\meta{path-to-inproof-reference-to-balanced-binary-tree}$"/>
</impacts>
\end{lstlisting}\vspace*{-3mm}
  \caption{Example Impact Annotation File}
  \label{fig:sampleimpacts}
\end{figure}

\paragraph{Propagation Phase of CIA.}\label{sec:cia:propagation}
The second, so-called \emph{propagation} phase, analyses the semantic graph and exploits the
information about semantics objects and relations being marked as
added, deleted or preserved to propagate the impact of changes through
the semantic graph. Impacts are a third type of nodes, different from the syntax and semantic nodes. They contain a human-oriented
description of the impact and can only be connected to semantic nodes. 
For instance, we have one marking a definition for some symbol, say
$f$, as being changed, when its body has changed. Furthermore, we have
rules that propagate that information further to definitions that
build upon $f$ or proofs using that definition (see
right-most rule of Fig.~\ref{fig:rulepropagate} for an example). Overall, we have 15
rules to analyse and propagate the impacts.


\paragraph{Projection Phase of CIA.}
Finally, we have the \emph{projection} phase which essentially
consists of one generic rule that projects the impact information of
the semantic nodes backwards along the origin links to the syntactic
node and creates a corresponding impact annotation for the syntactic
part of the documents.  The impact annotations are output in a
specific XML format, where an impact
annotation refers to the \lstinline[language=XML]{xml:id} of the
\OMDOC content element in its \lstinline[language=XML]{for} attribute
and the \lstinline[language=XML]{name} attribute contains the
human-oriented description of the impact. For our running example we obtain 
the impact shown in Fig.~\ref{fig:sampleimpacts}.

\subsection{Change Impact Analysis Workflow}\label{sec:cia:workflow}

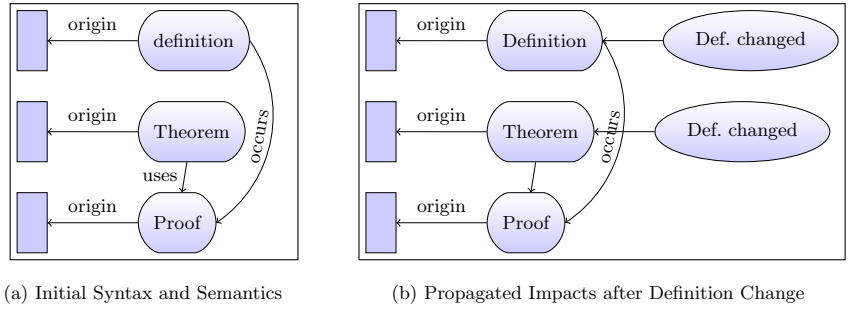
\begin{figure}[t]
  \centering

\tikzstyle{line} = [posline] 
\tikzstyle{propline} = [posline] 
\tikzstyle{rel} = [draw,->]

\begin{tikzpicture}[scale=.8,transform shape]





\node[
draw] (annotated){
\begin{tikzpicture}[node distance=0.5cm]
  \node [textnode] (def)  {};
  \node[block, right=1.5cm of def] (adef) {definition};

  \node [textnode, below=of def] (thm) {};
  \node[block, right=1.5cm of thm] (athm) {Theorem};

  \node [textnode, below=of thm] (uses) {};
  \node[block, right=1.5cm of uses] (auses) {Proof};

  \draw[line] (adef) -- node[midway,above] {origin} (def);
  \draw[line] (athm) -- node[midway,above] {origin} (thm);
  \draw[line] (auses) -- node[midway,above] {origin} (uses);

  \draw[rel] (athm) -- node[midway,left] {uses} (auses);
  \draw[rel] (adef.east) to [bend left=45] node[sloped,above,midway] {occurs} (auses.east);
\end{tikzpicture}
};

\node[below=.3cm of annotated,text width=50mm] {(a) Initial Syntax and Semantics};





\node[right=of annotated,draw] (changed2){
\begin{tikzpicture}[node distance=0.5cm]
  \node [textnode] (def)  {};
  \node[block, right=1.5cm of def] (adef) {Definition};
  \node[impact, right=1cm of adef] (idef) {Def. changed};
  \draw[propline] (idef) -- (adef);

  \node [textnode, below=of def] (thm) {};
  \node[block, right=1.5cm of thm] (athm) {Theorem};
  \node[impact, right=1cm of athm] (ithm) {Def. changed};
  \draw[propline] (ithm) -- (athm);

  \node [textnode, below=of thm] (uses) {};
  \node[block, right=1.5cm of uses] (auses) {Proof};

  \draw[line] (adef) -- node[midway,above] {origin} (def);
  \draw[line] (athm) -- node[midway,above] {origin} (thm);
  \draw[line] (auses) -- node[midway,above] {origin} (uses);

  \draw[propline] (athm) -- (auses);
  \draw[rel] (adef.east) to [bend left=45] node[sloped,above,midway] {occurs} (auses.east);
\end{tikzpicture}
};
\node[below=.3cm of changed2,text width=70mm] {(b) Propagated Impacts after Definition Change};






\end{tikzpicture}\vspace*{-3mm}
  \caption{Change Impact Analysis Phases}
  \label{fig:ciaphases}
\end{figure}

The workflow inside \doctip for the change impact analysis is to
initially build up a semantics graph for all documents that shall be
watched by \doctip. For our running example, the relevant parts of the
initial graph are given in Fig.~\ref{fig:ciaphases}(a). Upon a change
in some document, a semantic difference analysis between the old and
the new \OMDOC documents is performed, which results in a minimal
change description on an appropriate level of granularity.  This is
also provided by the \gmoc tool, which includes a generic semantic
tree difference analysis algorithm parameterized over document-type
specific similarity models.  The computed changes are applied to the
syntactic document graph. Subsequently, the impact analysis rule
systems of the three phases \emph{abstraction}, \emph{propagation},
and \emph{projection} are applied exhaustively in that
order.\footnote{Termination must be ensured by the designer of the
  rules systems. However, the \grgen-framework comes with a strategy
  language, that allows for a fine-grained control over the rule
  executions, which helps a lot for designing the strategies of the
  different phases. It is also used to sequentialize the three
  phases. } For our running example, we obtain a graph of the form in
Fig.~\ref{fig:ciaphases}(b). As a result of the impact analysis, the
\doctip system returns the computed impacts in the XML format
described before for \emph{all} documents it is maintaining (not only
for the document that caused the change).

\section{System Architecture}\label{sec:architecture}

\begin{figure}[t]
  \centering
  \usetikzlibrary{positioning,arrows,shapes,fit}

\tikzstyle{system} = [rectangle, draw, fill=blue!20, text width=1.6cm, text centered, rounded corners, minimum height=1cm,shade, top color=white, bottom color=blue!20 ]
\tikzstyle{arrowleft} = [single arrow, draw, inner sep=1pt, anchor=west,minimum width=1cm, minimum height=1.7cm, fill=lightgray!20,single arrow head indent=.4ex]
\tikzstyle{arrowright} = [single arrow, shape border rotate=180, draw, inner sep=1pt, anchor=west,minimum width=1cm,minimum height=1.7cm,fill=lightgray!20,single arrow head indent=.5ex]

\pgfdeclareimage[width=1cm]{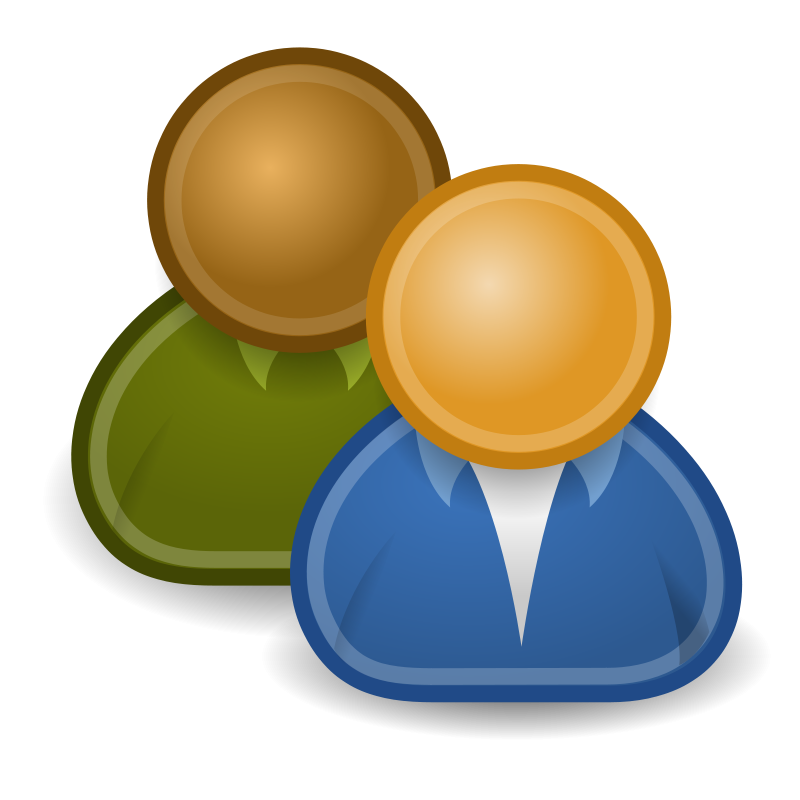}{user}

\begin{tikzpicture}[scale=0.75,transform shape]
  \node[rectangle] (user) {\pgfuseimage{user}};
  \node[system,right=1.7cm of user] (planetary) {\sys};
  \node[system,right=1.9cm of planetary] (tnt) {\tntbase};
  \node[system, right=1.9cm of tnt] (doctip) {\doctip};

  \node[above right=0.1cm of planetary, arrowleft] (ar1) {\stex};
  \node[below right=0.1cm of planetary, arrowright] (ar2) {XHTML};

  \node[above right=0.1cm of tnt, arrowleft] (ar3) {\OMDOC};
  \node[below right=0.1cm of tnt, arrowright] (ar4) {Impacts};
  \node [double arrow, left=0.1cm of planetary,draw, minimum height=1.7cm, minimum width=1cm, fill=lightgray!20, double arrow head indent=.5ex] {};
	
\node[rectangle,rounded corners,draw,inner sep=12pt,dashed, fit=(user) (doctip) (ar1) (ar4)] (pacbox) {};

\end{tikzpicture}
  \caption{Tool Chain}
  \label{fig:toolchain}
\end{figure}
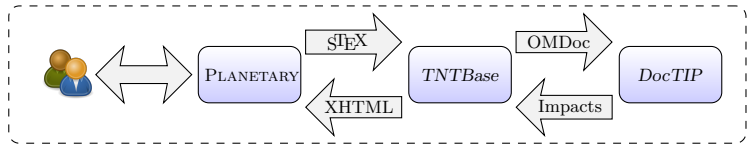

In order to add change management support for the workflows, we consider the architecture
of the \sys system (see Fig.~\ref{fig:toolchain}). The user interacts with the \sys system
via a web browser, which presents the mathematical knowledge items based on their
XHTML+MathML presentation in a WiKi-like form. The XHTML+MathML documents are rendered
from content oriented mathematical knowledge items in {\OMDOC} format. Along with the
XHTML+MathML document versions, the \sys system maintains the original \stex document
snippets, which the author can edit in the web browser. The {\OMDOC} documents are
maintained in the \tntbase repository together with their original \stex source
snippets. Any change in the \OMDOC documents in \tntbase results in an update of the
corresponding knowledge items in the \sys system after rendering the \OMDOC in
XHTML+MathML. Upon edit of the \stex snippets in the \sys system, a new \OMDOC is created
from the \stex sources~\cite{GinStaKoh:latexmldaemon11} and pushed into \tntbase, which
returns the XHTML+MathML presentation.

The \tntbase~\cite{ZhoKoh:tvsx09} is a Subversion based repository for normal files as well
as XML files. It behaves likes a normal Subversion repository, but offers special support
for XML documents by storing the revisions in a XML database. By this it allows efficient
access via XQueries to XML objects. \tntbase allows the definition of document specific
presentation routines, such as the XHTML+MathML rendering of \OMDOC documents. For its role
as repository for the \sys system, it is important to note that the \stex snippets and the corresponding \OMDOC documents are
stored together in the same directory in \tntbase, such as, for instance,

\begin{itemize}
\item the file \lstinline{balanced-binary-trees.tex} that contains the source of
  Fig.~\ref{fig:defs}, and
\item \lstinline{balanced-binary-trees.omdoc} that contains the \OMDOC transformation.
\end{itemize}

\noindent To add change management support, we connected \doctip to \tntbase. \doctip returns impact information in form of annotations to the \OMDOC documents, which are
stored in the \tntbase as an extra file together with the \OMDOC and the \stex files, but
with the extension ``.imp'', such as
\begin{itemize}
\item \lstinline{balanced-binary-trees.imp} (in the XML format shown in Fig.~\ref{fig:sampleimpacts}).
\end{itemize}

\noindent Like the change in the \OMDOC file, any change in the impacts file is forwarded as is by
\tntbase to the \sys system. The rendering of \OMDOC in XHTML+MathML preserves the
\lstinline[language=XML]{xml:id}. Therefore, the \sys assigns the impacts to the
XHTML+MathML snippets using the \lstinline[language=XML]{for}-attributes and presents on
the WiKi-page. 

\section{Example Revisited}\label{sec:reex}

\begin{figure}[t]
\includegraphics[width=\linewidth]{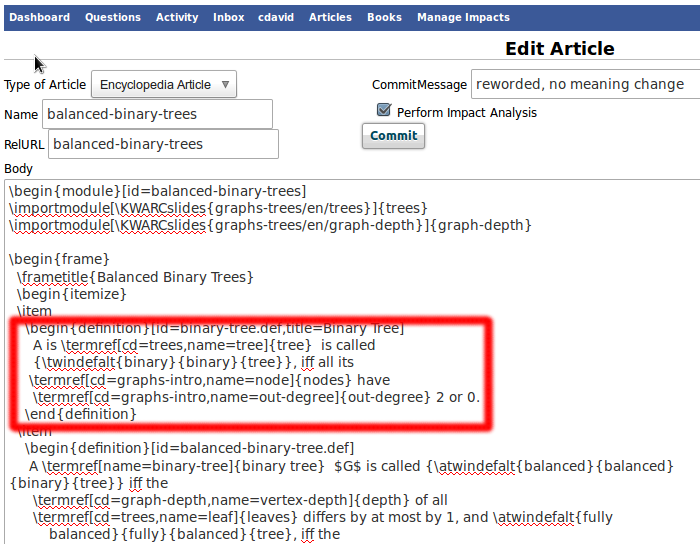}
\caption{Committing Changes  in \sys}\label{fig:commit}
\end{figure}
\begin{figure}[t]\centering
\includegraphics[width=\linewidth]{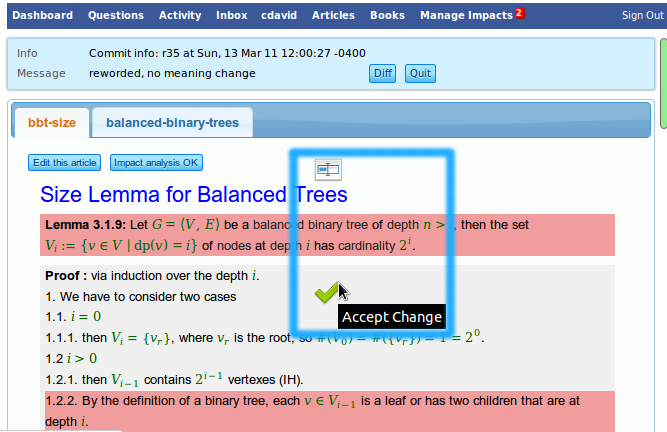}
\caption{The Impact Resolution Dialog}\label{fig:bbt-impact}
\end{figure}

To see how the parts of the system interact, let us revisit the
example from Section~\ref{sec:workflow}. Say the user found a typo in
the binary trees module in Fig.~\ref{fig:defs}. She opens the web
editor, corrects it and submits the changed module (see
Fig.~\ref{fig:commit}; note that the user requested a change impact
analysis). The system communicates the changes to \doctip, which
determines the list of impacts based on the semantic
relations. \doctip in turn communicates the impacts to \tntbase, which stores
them for further reference and passes them on to the \sys
system. Moreover, it notifies the user about impacts by updating the
superscript number on the ``Manage impacts'' field in the top menu bar
(see Fig.~\ref{fig:bbt-impact}). If the user decides to act on the
impacts, she gets the impact resolution dialog in
Fig.~\ref{fig:bbt-impact}, which has a tab for every module that is
impacted by the change. Note that the user gets the module in
its presented form as this is the most readable view, and,
furthermore, we can use the identifiers in the impacts (see
Fig.~\ref{fig:sampleimpacts}) to highlight the affected objects.  For
each of them, the user can then either discard the
impact information if it was a spurious impact (via the checkmark icon
in the ``Accept Change'' box) or edit the source of the impacted
object (via the ``edit'' icon in the box) and mark it as resolved
afterwards. Alternatively, she can use the action links above to make
changes at the level of the whole module. Note that a conventional
conflict resolution dialog via three-way merge as we know it from
revision control systems does not apply to this situation, since we
only have to deal with ``long-range conflicts'', i.e., impacts between
different objects, not conflicting changes to a single object.  When
the user quits the impact resolution dialog, all discarded and
resolved impacts are communicated to \tntbase together with the
changes.  \tntbase updates the set of tabled impacts and communicates the
changes to \doctip for a further round of CIA. Note that the storage
of tabled impacts in \tntbase (the additional ``.imp'' files) makes
the change management workflow more flexible over time. The need for
this was unanticipated before the integration and triggered a
re-design of the system functionality.

\section{Related Work}\label{sec:related}

There exists several methods for software development 
that estimate the scope and complexity of a change of a piece of
software with respect to other modules and documentation, known as
\emph{software change impact analysis}.  The methods are usually based on
modeling data, control, and component dependency relationships within
the set of source code. Such relationships can be automatically
extracted using well-known techniques such as \emph{data-flow
  analysis}~\cite{white92}, \emph{data dependency
  analysis}~\cite{krm88}, \emph{control flow analysis}~\cite{loyall93}
, \emph{program slicing}~\cite{korel90}, \emph{cross referencing} and
\emph{browsing}~\cite{bohner95}, and logic-based \emph{defects
  detection} and \emph{reverse engineering
  algorithms}~\cite{hwang98}. From an abstract point of view, we have
a similar set-up as we extract and collect relevant information and
their dependencies in the semantical extension of the document. For
example, the process of extracting dependencies between definitions,
axioms, and theorems and their uses in proofs can be seen to be
similar to a data-flow or dependency analysis for software. However,
on the concrete level, our approach differs because the flexiformal
documents we deal with do not have a formal semantics as software
artefacts. Indeed, we cannot directly interpret the textual parts of
\stex documents, but have to rely on the \stex markup manually
provided by the author. Thus, the impact analysis can always only be
as accurate as the manual annotations are. Furthermore, not having a
formal semantics at hand, we cannot automatically check if a certain
change really has an impact on other parts. In order to be
``complete'', we have to follow a possibilistic approach to propagate
impacts and thus may get false positives, i.e., spurious impacts. Since
impact information for some parts may trigger further impact
propagations (due to the possibilistic approach), this may result in
many spurious impacts in principle. For this a dependency management
on impacts nodes themselves can be used (by adding dependency links
between impacts in the rules) in order to propagate the deletion of
spurious impact information by the user.

Requirement tracing~\cite{Jarke98} is the process of recording
individual requirements, linking them to system elements, such as
source code, and tracing them over different levels of
refinement. Several tools have been developed to support requirement
tracing, such as the {\doors} system~\cite{doors:web}. Within our
setting, the change of an object, e.g., a definition, gives rise to an
impact, such as to revise the proof of a theorem.  Similar to
requirements, these impacts are linked to concrete objects and may
depend on each other. Also similar is that requirements are formulated
in natural language and the requirements tracing system has no access
to the semantics, hence also has to follow a possibilistic approach.
Of course, the type of relationships between requirements are tailored
to that domain in requirement tracing as they are in our case. The
main difference is that with our approach the relationships are not
built into the tool, but can be defined externally in separate rule
files. This allows the addition of new relationships, types of impacts and
propagation rules, for instance in order to accommodate the various
extensions of \OMDOC like for exercises, but also for didactic
knowledge. This will enable to add change impact analysis for
E-learning systems like ActiveMath~\cite{activemath:online}, that are based on \OMDOC with
their own didactic extensions and that lack change impact analysis
support for the authors of course materials.



\section{Conclusion}\label{sec:concl}

We have presented an integration of a management of change functionality into an active
document management system. The combined system uses the semantic relations that were
originally added to make documents interactive to propagate impacts of changes and
ultimately help authors keep the collections of source modules consistent. The approach is
based on impact analysis via graph rewriting rule systems for a core of the \OMDOC
format. CIA support for extensions of that core \OMDOC can easily be added on demand and,
due to the generic nature of impact descriptions and their handling in \sys, the
presentation module does not need to be adapted.

One limitation of the current integration that we want to alleviate in
the near future is that our integration currently assumes a
single-user mode of operation, as we have no means yet to consistently
merge the three kinds of documents (\stex, \OMDOC and impacts file).
Moreover, multiple users working on different branches that are partly
merged on demand are also not supported yet. One of the main
conceptual problems to be solved here is how to deal with propagating
changes by ``other authors''. For that we plan to build in the notion
of versioned links proposed in~\cite{KohKoh:micvl11}. Finally, the
current policy to eagerly trigger the change impact analysis after
each edit may be undesirable in situations where the author wants to
perform several small edits, which currently may result in many
spurious impacts. The impact analysis policy is pre-configured in
\doctip and we could easily enable the author to change it via the
\sys system. However, it requires a mechanism to enforce impact
analysis eventually, in order to prevent to just turn it off. For this
we need to gather more experience what would be a suitable policy to
optimally fit into the workflows.

\paragraph{Acknowledgements.} We would like to thank the anonymous
reviewers for their feedback and Andrea Kohlhase who semantically
annotated manually an extended document corpus for discussions about
the presented workflows and her comments on earlier versions of this
paper.

\printbibliography
\end{document}